\documentclass[twocolumn,english,aps,prb,twocolum,superscriptaddress,natbib,bibnotes,amsmath,amssymb,floatfix,groupedaddress,footinbib]{revtex4-2}
\usepackage[colorlinks=true,citecolor=blue,linkcolor=magenta]{hyperref}
\usepackage[addedmarkup=colored, authormarkupposition=left]{changes} 

\usepackage{soul}
\definechangesauthor[name={TJK}, color=red]{TJK}

\usepackage[english]{babel}
\babelprovide{en}
\usepackage{amsmath,amsfonts,amssymb}
\usepackage[T1]{fontenc}
\usepackage{xurl}
\usepackage{amsmath}
\usepackage{siunitx}
\usepackage[version=4]{mhchem}
\usepackage{amsfonts}
\usepackage{amssymb}
\usepackage{epstopdf}
\usepackage{graphicx}
\graphicspath{{./Figures/}}
\usepackage[utf8]{inputenc}
\UseRawInputEncoding

\newcommand{\ErSiN}{Er:Si$_3$N$_4$ }
\newcommand{\SiN}{Si$_3$N$_4$ }

\begin{document}
	
	\title{A fully integrated dispersion-managed femtosecond mode-locked laser}
	
	\author{
		Xurong Li$^{1,2*}$,
		Xuan Yang$^{1,2*}$,
		Zheru Qiu$^{1,2*}$,
		Jiale Sun$^{1,2*}$,
		Xinru Ji$^{1,2}$,
		Jianqi Hu$^{1,2}$,
		Grigorii Likhachev$^{1,2,3}$,
		Yichi Zhang$^{1,2}$,
		Ulrich Kentsch$^{4}$,
		and Tobias J. Kippenberg$^{1,2\dag}$}
	
	\affiliation{
		$^1$Institute of Physics, Swiss Federal Institute of Technology Lausanne (EPFL), CH-1015 Lausanne, Switzerland\\
		$^2$Institute of Electrical and Micro Engineering, Swiss Federal Institute of Technology Lausanne (EPFL), CH-1015 Lausanne, Switzerland\\
		$^3$EDWATEC SA, CH-1015 Lausanne, Switzerland\\
		$^4$Helmholtz-Zentrum Dresden-Rossendorf (HZDR), 01328 Dresden, Germany\\
		$*$: These authors contributed equally.
	}
	
	\maketitle
	\noindent\textbf{
		Femtosecond lasers underpin a wide range of applications, including tracking chemical reaction dynamics \cite{zewail_femtochemistry_2000}, material processing \cite{KannateyAsibu2023}, and corneal surgery \cite{soong_femtosecond_2009}. Their regular pulse trains form optical frequency combs that have revolutionized timekeeping \cite{beloy_frequency_2021}, spectroscopy \cite{picque_frequency_2019}, distance measurement \cite{coddington_rapid_2009}, and time transfer \cite{xin_attosecond_2017}, among numerous applications.
		On-chip optical frequency combs have been achieved, e.g., through parametric interactions in Kerr media \cite{Kippenberg2018}, and, by virtue of their high repetition rates, have expanded applications into areas such as optical communications \cite{rizzo_massively_2023}, microwave photonics \cite{kudelin_photonic_2024,Sun2024,zhao_all-optical_2024} and neuromorphic computing \cite{Feldmann2021}.  
		Yet, to date only femtosecond lasers, e.g., based on rare-earth-doped fibers, can access low repetition rates (i.e., from 100~MHz to 1~GHz) that promote high peak intensities, while integrated, chip-scale ultrafast sources today operate with repetition rates that are orders of magnitude higher, typically well beyond 10 GHz.  
		Here, we demonstrate a self-starting, photonic integrated mode-locked laser, based on a dispersion-managed mode-locking architecture, that accesses this parameter regime. The laser combines erbium-implanted \SiN gain waveguides, integrated chirped Bragg gratings, and a semiconductor saturable absorber mirror.
		It generates optical pulses with repetition rates from 0.5 to 1.2 GHz, pulse durations as short as 300 fs, and mode-locking thresholds down to 27.3 mW. 
		The output forms a passively stable optical frequency comb with a measured comb-line drift of less than 1\% of the repetition rate, surpassing the stability of commercial fiber-based mode-locked lasers by two orders of magnitude.
		Leveraging the ultra-low threshold of this platform, we achieve complete hybrid integration by co-packaging the laser with a telecom-grade 980-nm III-V pump diode chip inside a compact photonic module. 
		The resulting integrated electrical-in/optical-out module delivers turnkey, stable mode-locked pulses, offering a compact, low-power, and vibration-insensitive foundry-compatible platform for field-deployable applications in optical metrology and precision sensing.
	}

	
	Optical frequency combs (OFCs) have revolutionized numerous applications, including high-precision optical clocks \cite{beloy_frequency_2021}, ultra-stable microwave generation \cite{fortier_generation_2011}, spectroscopy \cite{picque_frequency_2019}, and astronomical spectrograph calibration \cite{steinmetz_laser_2008}. To date, the leading OFC platforms are fiber-based femtosecond mode-locked lasers (MLLs) with rare-earth gain, which have been used in nearly all OFC applications \cite{Diddams2020}. 
	Numerous passive mode-locking techniques to produce fiber-based femtosecond pulses have emerged, such as nonlinear loop mirror \cite{hansel_all_2017}, nonlinear polarization evolution \cite{hofer_mode_1991}, and semiconductor saturable absorber mirror (SESAM) \cite{Keller2010}.
	Typically, these lasers operate within the 100-MHz to 1-GHz repetition-rate regime \cite{byun_compact_2010,lesko_fully_2020}, yielding dense comb spectra ideal for high-resolution spectroscopy \cite{diddams_molecular_2007}. The large ratio of pulse interval and pulse duration also provides high peak powers required to drive nonlinear processes, such as supercontinuum generation \cite{dudley_supercontinuum_2006} or terahertz wave emission \cite{tripathi_fiber-laser_2013}.
	Translating these sources onto photonic integrated circuits (PICs) promises to expand OFC deployment in applications where size, weight, power consumption, and cost are critical, yet has remained elusive to date.

	Significant progress has been made in PIC-based frequency comb generators, most notably dissipative Kerr soliton (DKS) microcombs \cite{Kippenberg2018}.
	They typically produce frequency combs with repetition rates spanning from tens of GHz to THz regime, enabling numerous system-level applications, including massively scalable photonic data links \cite{rizzo_massively_2023}, astrophysical spectrometer calibration \cite{obrzud_microphotonic_2019},
	neuromorphic computing \cite{Feldmann2021} or low-noise microwave oscillators \cite{kudelin_photonic_2024,Sun2024,zhao_all-optical_2024}. 
	It would be desirable to also access the canonical, lower repetition rates with photonic integrated circuit-based frequency combs.
	For instance, high-resolution spectroscopy requires fine comb spacing to achieve Doppler-limited resolution for precision gas sensing \cite{picque_frequency_2019}.
	Furthermore, reducing the repetition rate is a key strategy for generating ultrashort pulses with high peak powers, which are necessary for nonlinear Raman spectroscopy \cite{Freudiger2014}, two-photon microscopy \cite{Xu2013}, terahertz wave emission \cite{tripathi_fiber-laser_2013}, supercontinuum generation \cite{Zia2023}, and self-referenced frequency combs \cite{Udem2002}.
	Specifically, self-referencing of a frequency comb requires octave-spanning supercontinuum generation—a process strictly gated by peak power, which is a function of both pulse energy and duration \cite{johnson_octave-spanning_2015}. To reach these nonlinear thresholds without exceeding the average power limits of integrated platforms, the repetition rate must be significantly reduced for high pulse energies ($\mathrm{E}_\mathrm{pulse}=\mathrm{P}_\mathrm{avg}/\mathrm{f}_\mathrm{rep}$), while short pulse durations further enhance peak powers ($\mathrm{P}_\mathrm{peak}\approx\mathrm{E}_\mathrm{pulse}/\mathrm{\tau}_\mathrm{pulse}$).
	
	However, accessing repetition rates in the $\lesssim$ 1-GHz regime on-chip remains a formidable challenge. 
	For DKS microcombs, reducing repetition rates leads to a sharp decline in pump-to-comb power conversion efficiency \cite{bao_nonlinear_2014} and an increased parametric oscillation threshold ($\sim$V$/$Q$^2$) due to the larger mode volumes and finite quality factors \cite{kippenberg_kerr-nonlinearity_2004}. 
	Similarly, for EO combs, lowering the repetition rate often compromises comb bandwidth, device footprint, and power consumption \cite{shams-ansari_low-repetition-rate_2020}.
	Furthermore, in high-confinement waveguides in PICs, the intense peak powers associated with low-repetition-rate pulses can trigger extreme Kerr nonlinearities, causing pulse destabilization and temporal break-up \cite{Anderson1992}. 
	
	The ideal solution would be to translate the success of rare-earth-doped fiber MLLs into the PIC domain. Rare-earth ions provide wide gain bandwidths necessary to support femtosecond pulse generation, while offering high gain and low noise due to their millisecond-long upper-state lifetimes \cite{BECKER1999xv}. Furthermore, they are temperature-insensitive, can be optically pumped by low-cost III-V pump lasers, and are compatible with ion implantation \cite{Liu2022}. Realizing such sources on-chip would not only inherit these material benefits but also unlock the transformative advantages of photonic integration, including wafer-scale manufacturability and the seamless incorporation of advanced functionalities like monolithic piezoelectric actuators \cite{liu_monolithic_2020}, high-speed on-chip modulators \cite{churaev_heterogeneously_2023}, or dual-comb architectures \cite{coddington_dual-comb_2016}.
	Pioneering demonstrations of erbium-doped silica waveguide MLLs marked an important milestone toward this goal \cite{Byun2009}. However, their weakly guiding nature results in large effective mode areas comparable to standard optical fibers, constraining integration density and demanding multi-centimeter-scale footprints. Furthermore, the massive mode-field diameter mismatch between these diffuse silica channels and the highly confined facets of III-V pump diode chips fundamentally precludes compact, chip-scale hybrid integration of the pump laser with the MLL cavity \cite{Liu2024}. 
	To reduce system footprints, recent efforts have shifted toward high-confinement, rare-earth-doped waveguide platforms. However, these architectures have largely been restricted to Q-switching \cite{Singh2024} or unstable Q-switched mode-locking \cite{Shtyrkova2019}. This limitation stems from excessive waveguide nonlinearities, where the combination of high nonlinear refractive indices and tight optical confinement triggers severe pulse breakup and instabilities \cite{Anderson1992}. 
	Other integrated technologies, including III-V semiconductor MLLs \cite{Hermans2022} and active MLLs on LiNbO$_3$ PICs \cite{Guo2023}, have similarly struggled to simultaneously achieve repetition rates in the $\lesssim$ 1-GHz regime and femtosecond pulse durations.
	Recently, a major breakthrough was achieved with the photonic integrated Mamyshev oscillator \cite{qiu_high-pulse-energy_2026}, which leverages self-phase modulation and spectral filtering to unlock stable mode-locking alongside nanojoule-level pulse energies. Despite this impressive performance, the high operational threshold renders it incompatible with direct, chip-scale pump integration \cite{Liu2024}. Because it relies on high-power external optical pumping, translating such systems into fully autonomous, self-contained micro-packages remains a formidable challenge.

	Here, we demonstrate a self-starting, photonic integrated circuit-based MLL (PIC-MLL) on an erbium-doped silicon nitride (Er:Si$_3$N$_4$) platform that overcomes the pulse-breaking limit via a dispersion-managed mode-locking architecture \cite{Turitsyn2012}. By incorporating integrated chirped Bragg gratings (CBGs) in \ErSiN waveguides \cite{Du2020}, we create a cavity where optical pulses periodically stretch and compress, reducing the effective nonlinearity and suppressing pulse-breaking. The PIC-MLLs operate at repetition rates of 0.5 to 1.2 GHz, a regime difficult to access in on-chip nonlinear OFC generators, and produce stable pulses as short as 300 fs. 
	The devices exhibit turnkey operation with mode-locking thresholds down to 27.3 mW, and feature exceptional comb-line stability that surpasses state-of-the-art commercial fiber lasers by two orders of magnitude. 
	With such power efficiency, we achieve a fully packaged, hybrid-integrated system by directly co-packaging the PIC-MLL with a 980-nm III-V pump diode chip inside a photonic module, providing a compact electrical-in/optical-out ultrafast source.

	\subsection*{Dispersion-managed cavity design and implementation}
	
\begin{figure*}[!h]
		\centering
		\includegraphics[width=\textwidth]{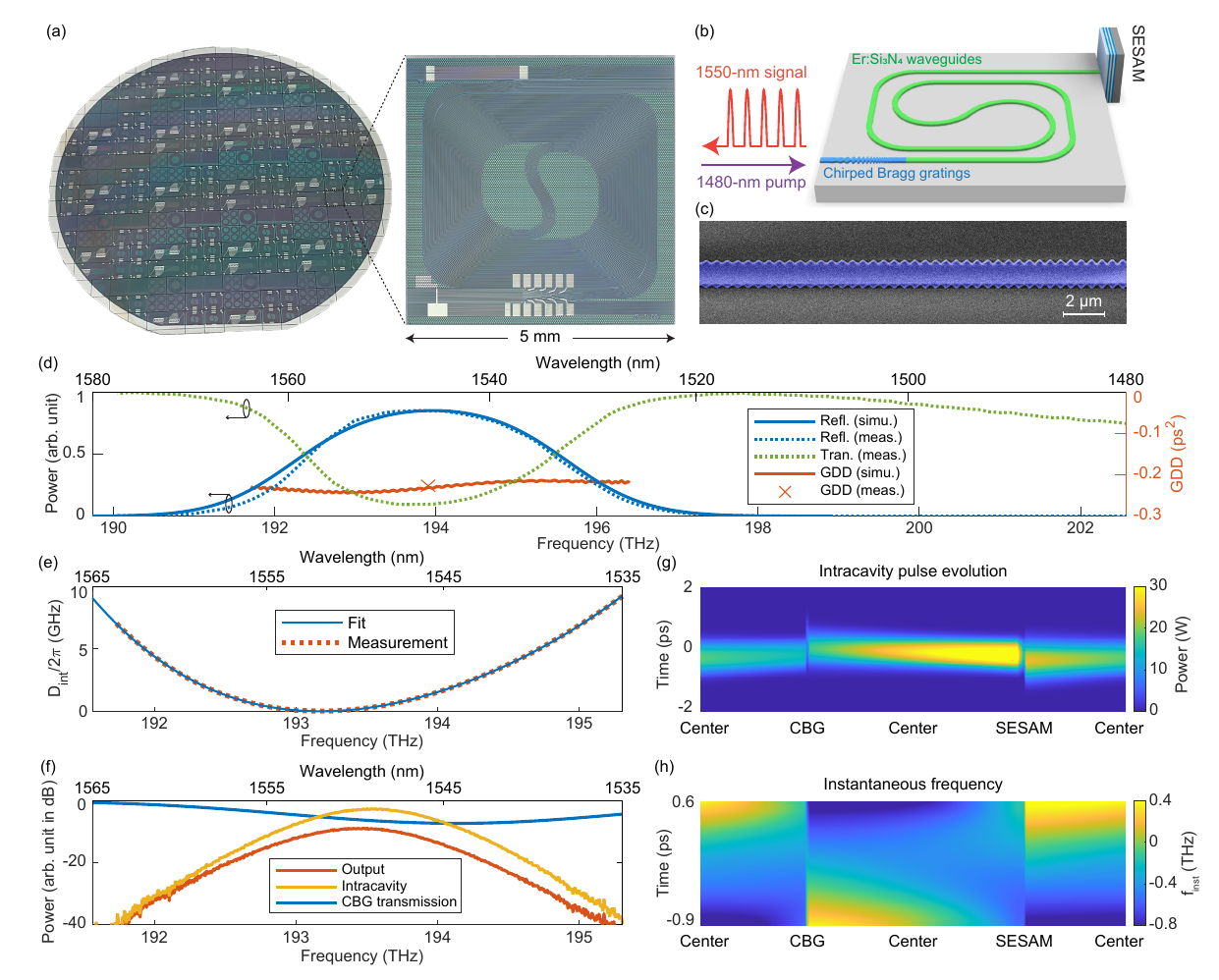}
		\caption{ 
			\footnotesize
			\textbf{Self-starting, photonic integrated femtosecond mode-locked lasers based on dispersion-managed architecture.} 
			\textbf{a,} Photograph of a 4-inch wafer of photonic integrated MLLs (left) and a $5\times5$ mm$^2$ chip containing 12 individual MLLs (right).
			\textbf{b,} Schematics of the photonic integrated MLL. Chirped Bragg gratings (CBGs) serve simultaneously as a cavity mirror, an output coupler, a pump injector, and a dispersion compensator. \ErSiN waveguides provide gain for the MLL. A SESAM butt-couples to the MLL chip as a mode-locking initiator. 
			\textbf{c,} False-colored scanning electron microscope image of a section of CBGs. 
			\textbf{d,} Simulated and measured spectra of CBGs' reflection, transmission, and group-delay dispersion (GDD). Both reflection and transmission spectra are normalized to the maximum of the transmission response. The CBG has 3000 periods, a center waveguide width of 1.2 $\mu$m, and a maximum waveguide width corrugation of 150 nm. The light incident from the short-period side of CBGs experiences negative GDD. 
			\textbf{e,} Measured and fitted integrated dispersion of the Fabry--P\'erot cavity, consisting of a pair of identical but opposite-oriented CBGs.
			\textbf{f,} Simulated intracavity and output optical pulse spectra, as well as the transmission response of CBGs.
			\textbf{g,h,} Intracavity pulse (\textbf{f}) and its instantaneous frequency (\textbf{g}) evolution in one cavity round trip during stable mode-locking. 
		}
		\label{Fig:1}
	\end{figure*}

	Our PIC-MLL is designed to exploit the interplay between on-chip gain, saturable absorption, and tailored dispersion (Fig.~\ref{Fig:1}b). The cavity comprises three primary functional blocks: \ErSiN gain waveguides, a butt-coupled SESAM for mode-locking initiation, and integrated CBGs. To overcome the pulse-breaking limit, we utilize a dispersion-managed scheme where the high-confinement gain waveguides provide normal dispersion, while the CBGs provide concentrated anomalous dispersion. This configuration forces the pulse to stretch and compress periodically, reducing the effective peak power and suppressing nonlinear instabilities. A SESAM was selected as the mode-locking initiator due to its commercial maturity and widely tunable parameters, such as modulation depth and recovery time \cite{Keller2010} (see Supplementary Information S3).
	
	We validated the dispersion-managed architecture using a numerical model based on the nonlinear Schr\"odinger equation, solved via the split-step Fourier method \cite{agrawal2000nonlinear} (see Supplementary Information S4). Our simulations capture the full laser dynamics, from the evolution of random noise to the steady-state mode-locked pulse within $\sim$100 round trips (see Supplementary Information S7). The results reveal the hallmark signatures of stretched-pulse operation: the instantaneous frequency chirp alternates in sign during each cavity round trip (Fig.~\ref{Fig:1}h), and the intracavity pulse exhibits a characteristic sech$^2$-like profile (Fig.~\ref{Fig:1}f). Significantly, the alternating-dispersion design suppresses phase-matched coupling to dispersive waves, resulting in very weak Kelly sidebands. 
	Our stability analysis further indicates that the device operates at a pulse energy exceeding the continuous-wave mode-locking stability threshold by a factor of 407, ensuring exceptional immunity to Q-switching instabilities \cite{Honninger1999}.
	
	The PIC-MLLs were fabricated on a commercially available 200-nm \SiN PIC platform using a wafer-scale erbium ion-implantation process \cite{ji_wafer-scale_2026} (see Supplementary Information S5). Each $5 \times 5$~mm$^2$ chip houses a parallel bundle of 12 MLLs with linear cavity lengths of $\sim$7 cm (Fig.~\ref{Fig:1}a), corresponding to repetition rates of 1.2 GHz. 
	The size of the individual chip can be further miniaturized to 0.5 mm $\times$ 2.5 mm, hosting only one MLL and enhancing the scalability of manufacture (see Supplementary Information S8L).
	The gain medium consists of \ErSiN single-mode waveguides with a $1.6~\mu$m $\times$ 200~nm cross-section, providing on-chip optical gain exceeding 1~dB/cm. 
	The CBGs serve simultaneously as a high-reflectivity mirror, an output coupler, a pump injector, and a dispersion compensator. To ensure stable mode-locking, the CBGs utilize Gaussian-apodized corrugations to minimize group-delay oscillations \cite{Matuschek1999}. 
	Measured group-delay dispersion (GDD) via an auxiliary on-chip Fabry--P\'erot cavity yielded a value of $-0.226$~ps$^2$ (Fig.~\ref{Fig:1}d) using frequency-comb-assisted spectroscopy \cite{liu_frequency-comb-assisted_2016}, in excellent agreement with our numerical designs and confirming the precise control over the cavity's net anomalous dispersion (see Supplementary Information S2).
	
	\subsection*{Mode-locked laser operation and characterization}
	
	Our PIC-MLL operates in a simple setup requiring only three components: a pump laser, a fiber-based wavelength-division multiplexer (WDM) , and a SESAM (Fig.~\ref{Fig:2}a; see Supplementary Information S6). The input and output share the same on-chip port, with the output collected through the WDM's 1550-nm port for characterization. The laser generates a stable pulse train without evidence of Q-switching (Fig.~\ref{Fig:2}c), and more than ten harmonics of the 1.2-GHz repetition rate are resolved on an electrical spectrum analyzer (Fig.~\ref{Fig:2}i). The optical spectrum shows a 1.14-THz (9.5-nm) 3-dB bandwidth centered at 1551~nm, corresponding to over 950 comb lines (Fig.~\ref{Fig:2}g,h). The PIC-MLL delivers 29.3~mW of on-chip power with 24.4-pJ pulse energy at the on-chip pump power of 221.1 mW (see Supplementary Information S8). Because the output pulses are strongly chirped, their transform-limited duration was retrieved using an intensity autocorrelator with a programmable pulse shaper for GDD control and an EDFA to compensate insertion loss (Fig.~\ref{Fig:2}d). 
	The dispersion compensation in the pulse shaper is equivalent to the linear pulse compression in optical fiber, and can be used to determine the optical fiber length for pulse dechirping.
	After compression, the shortest measured full-width at half-maximum (FWHM) duration was 424.7~fs, corresponding to a 300-fs Gaussian-type pulse (Fig.~\ref{Fig:2}e,f).

	\subsection*{Ultra-low-threshold and hybrid-integrated mode-locked lasers}
	
\begin{figure*}
		\centering
		\includegraphics[width=\textwidth]{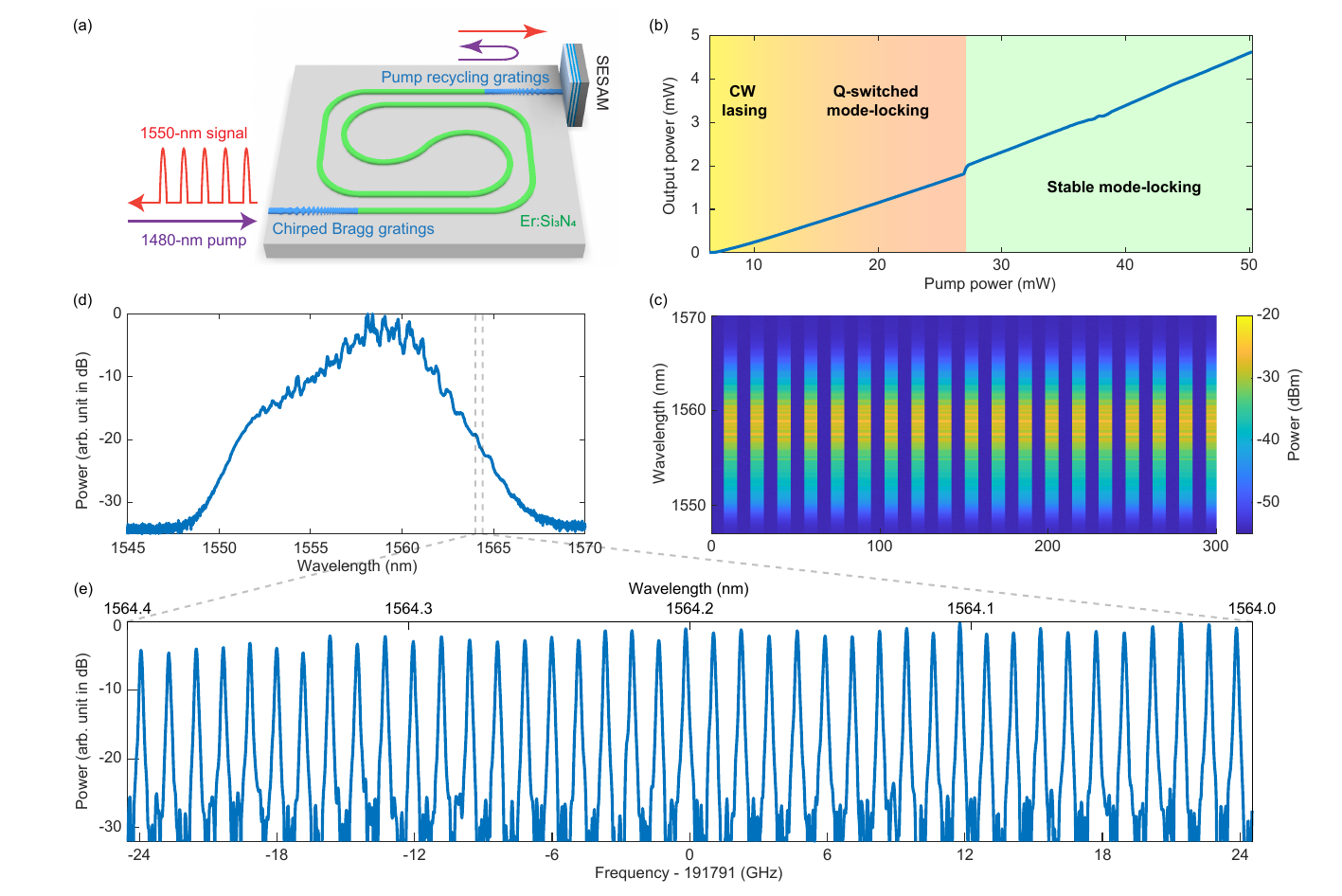}
		\caption{ 
			\footnotesize
			\textbf{Ultra-low-threshold photonic integrated mode-locked laser.} 
			\textbf{a,} Schematics of the ultra-low-threshold MLL with pump-recycling gratings, which reflect the residual pump light back to \ErSiN waveguides and transmit signal light toward SESAM. 
			\textbf{b,} On-chip output power of the MLL as a function of on-chip pump power during pump power ramping up. The MLL enters stable mode-locking state at a 27.3-mW pump power. 
			\textbf{c,} Measured optical spectrogram of the turnkey performance over nineteen consecutive switching tests.
			\textbf{d,} Normalized optical spectrum of the MLL measured by a regular OSA. 
			\textbf{e,} Normalized optical spectrum of the MLL measured by a high-resolution OSA (WaveAnalyzer 1500s), revealing comb lines with a 1.2-GHz frequency spacing. 
			Device ID: \texttt{D215\_02\_F8\_C8}.}
		\label{Fig:3}
	\end{figure*}

	We achieve an ultra-low pump threshold for stable mode-locking by incorporating pump-recycling gratings into the PIC-MLL architecture (Fig.~\ref{Fig:3}a). Designed by the same CBG principle, these pump-recycling gratings provide high reflectivity at the pump wavelength (1480 nm) while ensuring minimal distortion of the 1550-nm signals (see Supplementary Information S2). They reduce the mode-locking threshold through two mechanisms: (i) redistributing the 1480-nm pump power along the \ErSiN waveguides to enhance erbium ion inversion and optical gain; and (ii) suppressing residual 1480-nm light incident on the SESAM, thereby preventing continuous depletion of its modulation depth for phase-locking the 1550-nm pulses. With this approach, stable mode-locking is initiated at an on-chip pump power as low as 27.3 mW (Fig.~\ref{Fig:3}b), exhibiting robust turnkey operation over 19 consecutive power-switching cycles (Fig.~\ref{Fig:3}c).
	
	This ultra-low mode-locking threshold redefines the boundaries of system-level integration for ultrafast sources, enabling two operational paradigms. The first approach leverages remote optical pumping, where the PIC-MLL and SESAM are co-packaged in a photonic housing, and pump light is delivered via an optical fiber. As a  proof-of-concept prototype, we developed a portable system occupying a compact $15 \times 15$ cm$^2$ breadboard footprint. This setup integrates the packaged MLL, a commercial pump diode, and a control driver, and can be powered by just four standard AA batteries (see Supplementary Information S9).
	
\begin{figure*}
		\centering
		\includegraphics[width=\textwidth]{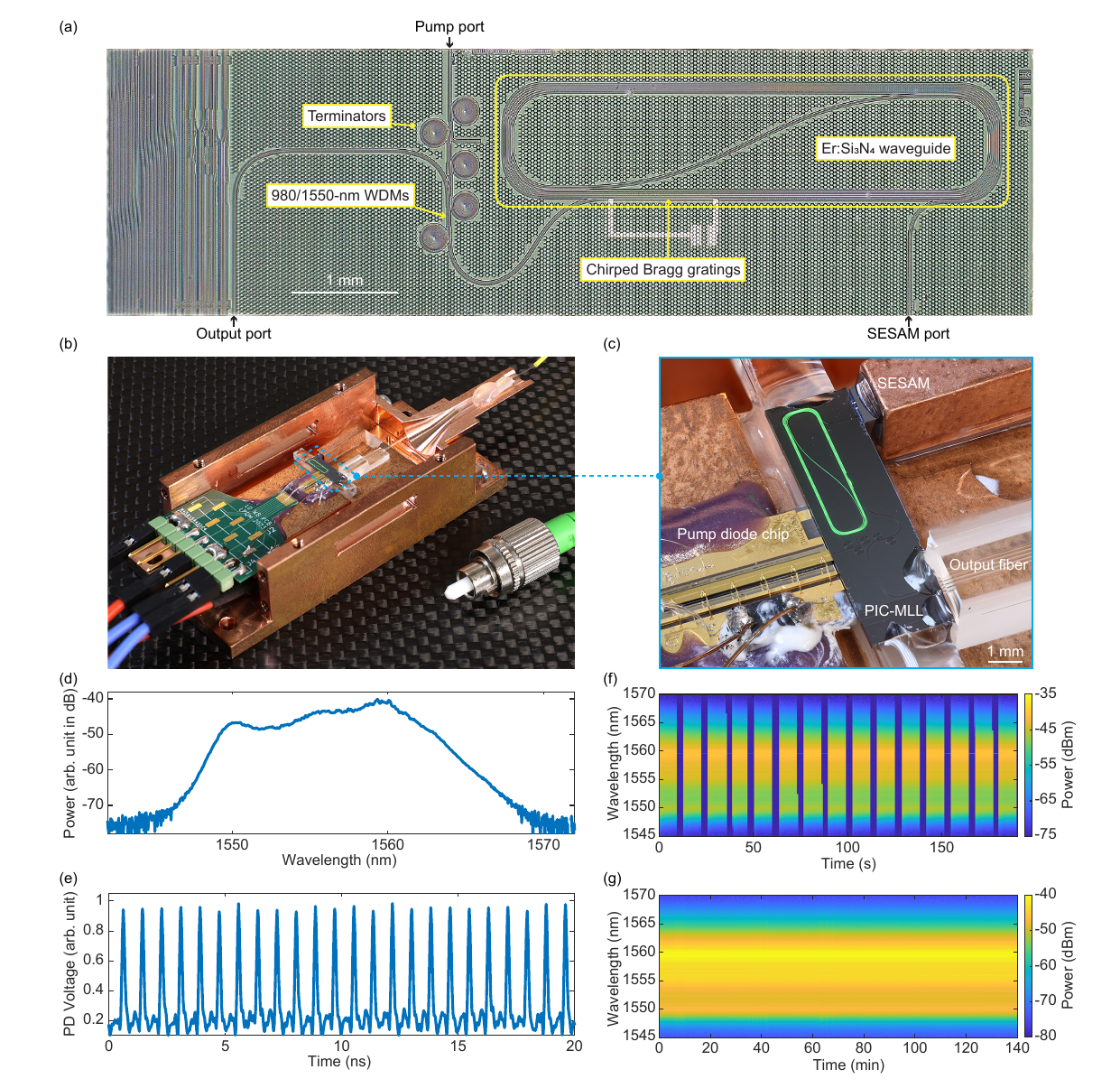}
		\caption{ 
			\footnotesize
			\textbf{Fully packaged hybrid-integrated PIC-MLL system and performance characterization.} 
			The hybrid-integration platform is configured as a three-port architecture. A 980-nm III-V pump diode chip is butt-coupled to the pump port for direct electrical-to-optical power conversion. A SESAM is integrated at the second port to initiate passive mode-locking, while the third port is pigtail-coupled to a single-mode fiber for optical output extraction. 
			\textbf{a,} Photograph of the standalone PIC-MLL chip prior to packaging.
			\textbf{b,} Photograph of the fully packaged PIC-MLL system within a photonic module.
			\textbf{c,} Close view of the hybrid integration.
			\textbf{d,} Output optical spectrum of the hybrid-integrated PIC-MLL. 
			\textbf{e,} Steady-state temporal pulse train recorded on an oscilloscope. 
			\textbf{f,} Measured spectrogram spanning 15 consecutive turnkey power-switching cycles. 
			\textbf{g,} Output optical spectrum continuously monitored over a 140-minute period.
			Device ID: \texttt{D252\_04\_F9\_C4}.
		}
		\label{Fig:hybrid_integration}
	\end{figure*}

	Furthermore, the exceptional power efficiency of this platform enables hybrid integration of the pump laser with the PIC-MLL, entirely eliminating external optical infrastructure and realizing a fully self-contained, electrical-in/optical-out module. A 980-nm III-V pump diode chip is directly butt-coupled to the pump port of the PIC-MLL. To prevent cavity destabilization, we engineered a cascaded four-stage 980/1550-nm integrated WDM that ensures highly efficient 980-nm pump injection while suppressing 1550-nm signal back-reflection at the diode emission facet (see Supplementary Information S9). Compared to 1480-nm excitation, the 980-nm pump provides higher gain efficiency in \ErSiN waveguides. This enables robust population inversion even in a single-pass configuration, while simultaneously capitalizing on mature, telecom-grade diode technology. 
	The pump diode chip, PIC-MLL chip, SESAM, and a single-mode output fiber are co-packaged within a compact photonic module (Fig.~\ref{Fig:hybrid_integration}a--c). Driven purely by electrical power, the fully integrated MLL exhibits highly reproducible, turnkey mode-locking (Fig.~\ref{Fig:hybrid_integration}d--f) and maintains continuous operation for over 140 minutes (Fig.~\ref{Fig:hybrid_integration}g). This operational autonomy demonstrates the compactness and field-readiness of our PIC-MLL architecture.

	\subsection*{Passively stable optical frequency comb} 
	
\begin{figure*}
		\centering
		\includegraphics[width=\textwidth]{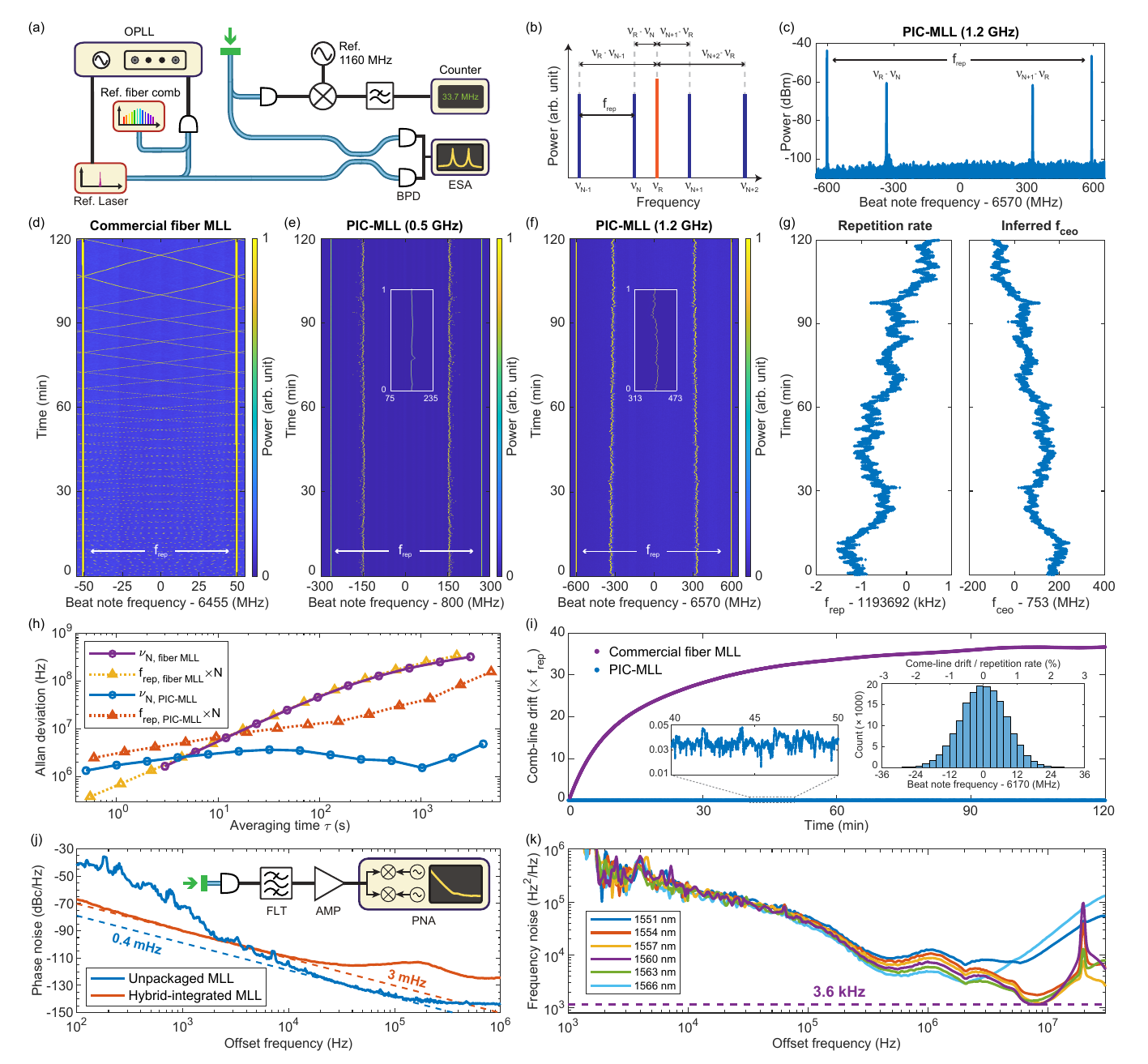}
		\caption{ 
			\footnotesize
			\textbf{Passively stable optical frequency comb generated by PIC-MLLs.} 
			\textbf{a,} Experimental setup for optical heterodyne beat note and repetition rate measurements of the PIC-MLL. 
			The reference laser (Toptica CTL) is phase-locked to a fully stabilized optical frequency comb (Menlo FC1500-ULNnova). 
			OPLL: optical phase-locked loop; BPD: balanced photodetector.
			\textbf{b,} Schematic illustration of the heterodyne beat note.
			\textbf{c,} Measured spectrum of the heterodyne beat note of the PIC-MLL at 1559.424 nm. 
			\textbf{d}--\textbf{f}, Measured heterodyne beat note spectrograms over 2 hours for a commercial fiber MLL (Menlo ELMO) (\textbf{d}), a 0.5-GHz PIC-MLL (\textbf{e}), and a 1.2-GHz PIC-MLL (\textbf{f}) against the reference laser. Insets display magnified 1-minute spectrograms. 
			\textbf{g,} Simultaneously recorded PIC-MLL's repetition rate (at 1.2 GHz) and the inferred f$_\mathrm{ceo}$. We assume the reference laser's frequency is closer to the left comb line.
			\textbf{h,} Allan deviation of the comb-line drift and comb-line-number-scaled repetition rate of both PIC-MLL and commercial fiber MLL. 
			\textbf{i,} Total comb-line drift. Left inset: zoomed-in plot for the 1.2-GHz PIC-MLL. Right inset: histogram distribution of PIC-MLL's comb-line drift over 2 hours. 
			The standard deviation is 8.0 MHz, corresponding to only $0.67\%$ of the repetition rate, demonstrating the exceptional passive stability of the frequency comb. 
			\textbf{j,} Phase noise of repetition-rate signals of an unpackaged and a hybrid-integrated PIC-MLL (at 1.2 GHz), with reference lines corresponding to 0.4-mHz and 3-mHz fundamental linewidths. 
			Inset: schematics of the experimental setup. FLT: filter; AMP: amplifier; PNA: phase-noise analyzer. 
			\textbf{k,} Frequency noise of heterodyne beat notes at 1.2-GHz PIC-MLL's different comb lines, indicating a 3.6-kHz intrinsic linewidth.
			Device ID: \texttt{D215\_02\_F1\_C8}, \texttt{D215\_02\_F8\_C8}, \texttt{D249\_04\_F4\_C2} and \texttt{D252\_04\_F9\_C4}.}
		\label{Fig:4}
	\end{figure*}

	The femtosecond pulse train of the MLL corresponds to a frequency comb in the spectral domain. Individual comb teeth are directly resolved on a high-resolution optical spectrum analyzer (Fig.~\ref{Fig:3}e), exhibiting a frequency spacing that precisely matches the 1.2-GHz repetition rate. The comb nature is further validated through heterodyne beating with a continuous-wave reference laser phase-locked to a fully stabilized optical frequency comb (Fig.~\ref{Fig:4}a--c). The PIC-MLL exhibits exceptional passive stability of its comb lines: during a continuous 2-hour heterodyne measurement, the standard deviation of the comb-line drift remained below 1\% of the repetition rate (Fig.~\ref{Fig:4}f,i). 
	This passive stability is an intrinsic feature of the photonic integration rather than an artifact of a specific cavity length, as evidenced by the identical comb-line stability measured in a 0.5-GHz PIC-MLL (Fig.~\ref{Fig:4}e).

	In stark contrast, a high-end commercial fiber MLL (Menlo ELMO) drifted by more than 33 times its repetition rate within just 1 hour under identical ambient conditions, and more than 8 times its repetition rate after the 30-minute warm-up time recommended by the laser manufacturer (Fig.~\ref{Fig:4}d,i). By simultaneously recording the PIC-MLL's repetition rate, we can directly infer the drift of the carrier-envelope offset frequency ($f_{\text{ceo}}$), which remained below one-third of the repetition rate throughout the 2-hour measurement window. Allan deviation analysis further confirms the superior comb-line stability of our integrated platform, showcasing a comb-line drift more than two orders of magnitude lower than that of the commercial fiber MLL at an averaging time of 1000~s (Fig.~\ref{Fig:4}h). 
	Fiber-based MLLs typically have no locking mechanism to stabilize the comb lines \cite{kim_ultralow-noise_2016}.
	For PIC-MLLs, the exceptional comb-line stability arises from the fixed peak reflection frequency of the on-chip CBGs, analogous to fiber Bragg grating-stabilized lasers \cite{Kashyap2010FiberBraggGratings}.
	
	The PIC-MLL also exhibits narrow and uniform optical linewidths. Frequency-noise characterization at six distinct wavelengths reveals variations of just 8.6\% and 12.3\% at 10-kHz and 100-kHz frequency offsets, respectively, spanning over 1,500 individual comb lines (Fig.~\ref{Fig:4}k). By fitting the noise floor near an 8-MHz offset, we estimate an intrinsic Lorentzian linewidth of 3.6~kHz. Furthermore, the single-sideband (SSB) phase noise of the photodetected signal at the 1.2-GHz fundamental repetition rate drops to $-$113.5~dBc/Hz at a 10-kHz offset (Fig.~\ref{Fig:4}j), yielding an integrated timing jitter of 46.9~fs (integrated from 10~kHz to 10~MHz). The extracted fundamental linewidth of the repetition-rate signal is 0.4~mHz (extrapolated from the 10--100~kHz). This represents a three-order-of-magnitude reduction over state-of-the-art heterogeneously integrated III-V MLLs \cite{Cuyvers2021} and stands on par with the fundamental linewidth achieved by the high-power photonic integrated Mamyshev oscillator \cite{qiu_high-pulse-energy_2026}.

	\subsection*{Discussion}
	
	We have demonstrated a self-starting, hybrid-integrated femtosecond mode-locked laser based on an \ErSiN PIC platform. By adopting a dispersion-managed architecture, our device overcomes the fundamental pulse-breaking limits inherent to high-confinement waveguides, enabling stable mode-locking on a photonic chip. The measured comb-line stability is two orders of magnitude superior to that of a high-end commercial fiber MLL, proving that integrated MLLs can outperform benchtop standards in passive stability. Furthermore, our PIC-MLL achieves a level of coherence surpassing all reported integrated semiconductor MLLs, while simultaneously delivering femtosecond pulses in the 1-GHz repetition rate regime (see Supplementary Information S1 for a comprehensive comparison). 
	Currently, the repetition rate stability is still limited by Gordon-Haus jitter—the coupling of center-frequency fluctuations to timing jitter within the net-anomalous dispersion regime \cite{kim_ultralow-noise_2016}. To reach the next tier of stability, future designs of this architecture can be dispersion-engineered closer to the near-zero-dispersion regime. This shift will not only suppress Gordon-Haus coupling but also support broader optical bandwidths and shorter pulse durations.
	
	The ultra-low mode-locking threshold of our PIC-MLLs redefines the possibilities for system-level integration. While alternative state-of-the-art architectures, such as integrated Mamyshev oscillators, offer impressive pulse energies, their high mode-locking thresholds require bulky, high-power external pumping systems that constrain their operation on optical tables \cite{qiu_high-pulse-energy_2026}. In contrast, the low-power operation of our dispersion-managed cavity enables hybrid integration with commercial 980-nm III-V pump diode chips.
	This advances the PIC-MLL from a passive photonic chip reliant on external optical infrastructure into a fully autonomous, turnkey, and field-deployable module. 
	
	Looking forward, the physical architecture demonstrated here is highly scalable and compatible with advanced volume manufacturing. 
	While SESAMs are currently butt-coupled, in-plane integration is feasible through grating couplers \cite{marchetti_coupling_2019}, supporting future approaches such as transfer printing \cite{Cuyvers2021} or direct wafer bonding \cite{lang_silicate_2021}.
	Alternatively, on-chip artificial saturable absorbers, such as nonlinear amplifying loop mirrors \cite{hansel_all_2017} and nonlinear interferometers \cite{Shtyrkova2019}, can be also adapted within this architecture to initiate passive mode-locking, enabling fully monolithic cavity designs.
	Furthermore, integrating monolithic piezoelectric actuators directly above the waveguide cavity and gratings \cite{liu_monolithic_2020} will facilitate active stabilization of both the repetition rate and carrier-envelope offset frequency.
	Finally, because this dispersion-managed scheme is wavelength-agnostic, it can be readily extended to other rare-earth-doped PIC platforms, such as ytterbium (1 $\mu$m), neodymium (1.33 $\mu$m), and thulium (2 $\mu$m).
	
	By successfully unifying turnkey femtosecond pulse generation, gigahertz repetition rates, and ultra-low thresholds with exceptional passive comb stability, wafer-scale manufacturability, and fully hybrid pump integration, this platform is uniquely positioned to transition benchtop ultrafast lasers into real-world deployments. We anticipate that these self-contained electrical-in/optical-out ultrafast modules will catalyze widespread adoption across diverse domains, including LiDAR sensing \cite{coddington_rapid_2009}, high-speed dual-comb spectroscopy \cite{coddington_dual-comb_2016}, portable terahertz systems \cite{li2023high}, and next-generation optical atomic clocks \cite{newman2019architecture}.

	\pretolerance=0
	
	\begin{footnotesize}
		
		
		\bigskip
		\noindent \textbf{Funding Information}: This work was supported by funding from the Swiss National Science Foundation under grant agreement No. 216493 (HEROIC), as well as the Horizon Europe EIC transition programme under grant No. 101291179 (PI-MOLL), and from the Air Force Office of Scientific Research under award No. FA9550-25-1-0259.
		
		\noindent \textbf{Acknowledgments}:
		The passive Si$_3$N$_4$ PIC fabrication and post-implantation process were conducted in the EPFL Center of MicroNanoTechnology (CMi). The ion implantation was performed at Helmholtz-Zentrum Dresden-Rossendorf (HZDR).
		
		\noindent \textbf{Author contribution}:
		X.L. and T.J.K. conceived the idea and concept. X.L., Z.Q., X.Y, and J.H. performed numerical simulation. X.L. designed the D215 MLL samples. X.Y., J.S., X.L., and Z.Q. designed the D244 and D252 reticles. X.J., X.Y., X.L, Z.Q., and Y.Z. fabricated the devices. U.K. and G.L. performed the ion implantation. X.L, Z.Q., X.Y., J.S., and J.H. performed the experiment and processed the data. J.S. and X.L. packaged the MLL. X.L. wrote the manuscript with contributions from all coauthors. T.J.K. supervised the work.
		
		\noindent \textbf{Data Availability Statement}: The code and data used in this work will be provided in a \texttt{Zenodo} repository upon publication of this manuscript.
		
		\noindent \textbf{Competing interests}: T.J.K. is a cofounder and shareholder of LiGenTec SA, a start-up company offering Si$_3$N$_4$ photonic integrated circuits as a foundry service. T.J.K. is a cofounder and shareholder of EDWATEC SA, a start-up company offering optical amplifiers on chip.
		
	\end{footnotesize}
	
	\bibliographystyle{apsrev4-2}
	\bibliography{bibliography_v2}

	\begin{figure*}
		\centering
		\includegraphics[width=\textwidth]{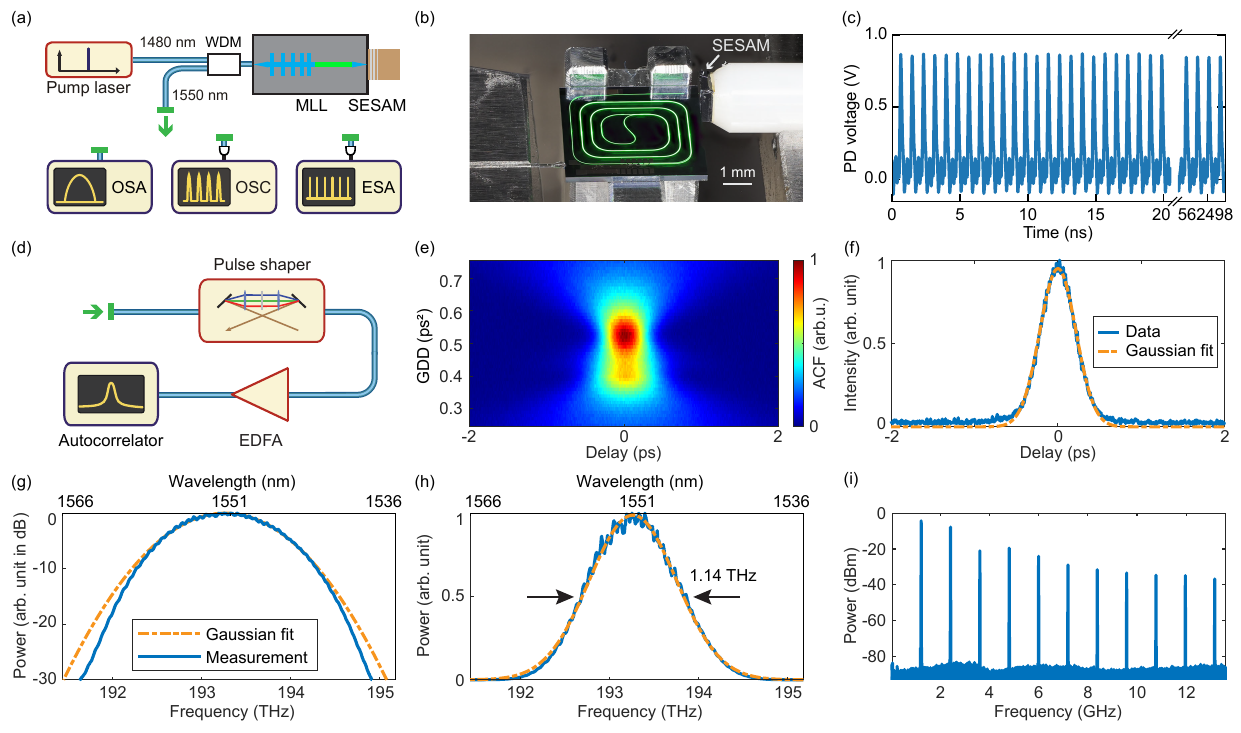}
		\caption{ 
			\footnotesize
			\textbf{Characterization of a photonic integrated dispersion-managed mode-locked laser.} 
			\textbf{a,} Schematic diagram of the experimental setup for MLL operation and characterization. OSA: optical spectrum analyzer; OSC: oscilloscope; ESA: electrical spectrum analyzer.
			\textbf{b,} Photograph of a MLL in operation.
			\textbf{c,} Stable pulse train measured by a 5-GHz photodetector and an OSC at a 376-mW on-chip pump power. 
			\textbf{d,} Schematic diagram of the experimental setup for measuring the transform-limited pulse width. 
			\textbf{e,} Autocorrelation function (ACF) as a function of GDD provided by the pulse shaper.
			\textbf{f,} Autocorrelation function at the shortest pulse width. The measured 424.7-fs FWHM pulse width of a Gaussian-like pulse in autocorrelation indicates a 300-fs pulse width.
			\textbf{g,h,} Logarithmic- and linear-scale optical spectra of the MLL output. The on-chip output and pump powers are 22.5 mW and 376 mW, respectively. 
			\textbf{i,} Microwave spectrum of the photodetected signal measured by an ESA, showing multiple harmonics of the repetition rate (1.2 GHz). The resolution bandwidth is 100 kHz, and the video bandwidth is 10 kHz.
			Device ID: \texttt{D215\_02\_F1\_C1} and \texttt{D215\_02\_F2\_C1}. 
		}
		\label{Fig:2}
	\end{figure*}
	
\end{document}